\documentclass[nofootinbib,aps,twocolumn,superscriptaddress]{revtex4}
\usepackage{amsmath}
\usepackage{citesort}
\usepackage{epsfig}

\newcommand{\beqa}{\begin{eqnarray}}
\newcommand{\eeqa}{\end{eqnarray}}
\newcommand{\beq}{\begin{equation}}
\newcommand{\eeq}{\end{equation}}
\newcommand{\bal}{\begin{align}}
\newcommand{\eal}{\end{align}}
\renewcommand{\Re}{{\cal R}e}
\renewcommand{\Im}{{\cal I}m}

\def\sla{\negthinspace\not\negmedspace}
\def\gsim{\ \rlap{\raise 3pt \hbox{$>$}}{\lower 3pt \hbox{$\sim$}}\ }
\def\lsim{\ \rlap{\raise 3pt \hbox{$<$}}{\lower 3pt \hbox{$\sim$}}\ }

\begin{document}

\preprint{BNL-HET-05/26}
\preprint{CMU-TH-05-09}
%\hbox{}
%\hbox{hep-ph/0510325}
%\hbox{September 2005}
%}}

\vspace*{18pt}

\title{Semi-inclusive hadronic $B$ decays as null tests of the Standard Model}

\def\addbnl{Physics Department, Brookhaven National Laboratory, Upton, New York 11973}
\def\addcmu{Department of Physics, Carnegie Mellon University, Pittsburgh, PA 15213}
\def\addIJS{J.~Stefan Institute, Jamova 39, P.O. Box 3000, 1001
Ljubljana, Slovenia}

\author{A. Soni}\affiliation{\addbnl} 
\author{J. Zupan} \affiliation{\addcmu}\affiliation{\addIJS}

\begin{abstract} \vspace*{18pt}
We propose a new set of observables that can be used as experimental null tests of the Standard Model in charged and neutral $B$ decays. The CP asymmetries in hadronic decays of charged B mesons into inclusive final states containing  at least one of the following mesons:
 $K_{S,L}$, $\eta'$, $c\bar c$ bound states or neutral $K^*$ or
 $D$ mesons, for all of  which a U-spin rotation is equivalent to a CP conjugation,
are CKM suppressed and furthermore 
vanish in the exact U-spin limit. We show how this reduces the theoretical error by using Soft Collinear Effective Theory  to calculate the CP asymmetries  for $K_{S,L} X_{s+d}$, $K^* X_{s+d}$ and $\eta' X_{s+d}$ final states in the endpoint region.
For these CP asymmetries only the flavor and not the charge of the decaying $B$ meson needs to be tagged up to corrections of NLO in $1/m_b$, making the measurements more accessible experimentally. 
\end{abstract}

\maketitle

%%%%%%%%%%%%%%%%%%%%%%%%%%%%%%%%%
%%%%%%%%%%%%% I %%%%%%%%%%%%%
%%%%%%%%%%%%%%%%%%%%%%%%%%%%%%%%%
\section{Introduction} 
Recently the experiments at the two asymmetric B-factories
have helped us attain an important milestone in our understanding of
CP violation phenomena. The Standard Model
(SM) prediction of $\sin 2 \beta = 0.742^{+0.072}_{-0.026}$ \cite{Charles:2004jd} is found to be 
in very good
agreement with the directly measured value 
$0.674 \pm0.026$ \cite{Hazumi}.
The effects of a CP-odd phase due to beyond the SM
sources are thus expected to cause only a small perturbation. Consequently {\it null tests} of
the SM gain special importance in our quest for new physics (NP).
Since CP is not an exact symmetry of the SM, it is
generally not possible to construct exact null tests
of the Cabibbo-Kobayashi-Maskawa (CKM) paradigm of CP
violation~\cite{ckm}; the best we can hope for are approximate
null tests.
One such null test
that has attracted a lot of attention lately is the
prediction that the difference of $S$ parameters in time dependent $B(t)\to J/\Psi K_{S,L}$ and ``penguin-dominated" 
decays such as 
$B(t)\to (\phi, \eta') K_{S,L}$ should be well below $5\%$~\cite{gw,ls,Beneke:2005pu,Williamson:2006hb,Cheng:2004ru,SU3}, 
which is violated at present
by about $1-2 \sigma$~\cite{HFAGSummer2006}. 
In this paper we propose a new set of observables
that can be used as rather clean and stringent null tests of the SM.

The proposed tests involve {\it direct} CP violating partial width differences (PWD) of {\it untagged} semi-inclusive
hadronic  decays of charged B mesons
\beq\label{CPincl}
\Gamma(B^-\to M^0 X_{s+d}^-)-\Gamma(B^+\to \bar M^0 X_{s+d}^+),
\eeq
with the notation that 
$M^0 X$ is a final state containing {\it at least one} meson $M^0$. For judicial choice of meson $M^0$
the PWD
in Eq. \eqref{CPincl} is doubly suppressed. In addition to the CKM suppression to be discussed in more detail below, 
it also vanishes in the limit of exact U-spin, if meson $M^0$ is either (i) an eigenstate of
discrete transformation $s\leftrightarrow d$, %for instance if it is 
such as $K_{S,L}$,
 $\eta'$ or any $c \bar c$ bound state, or (ii) if $M^0$ and $\bar M^0$ are related through $s\leftrightarrow d$ 
 transformation, e.g. $M^0$ can be $K^{0}$, $K^{0*}$ or $D^0$. In this latter case a sum over the 
two related states
 needs to be made, e.g. $\Gamma(B^-\to M^0 X_{s+d}^-)\to \Gamma(B^-\to (K^{0*}+\overline{ K^{0*}}) X_{s+d}^-)$.
Because of the double suppression SM predicts vanishingly small ({\it i.e.} $<1\%$) asymmetries
in these modes, while theoretical uncertainties on the predictions are reduced due to approximate SU(3) symmetry, so that they constitute useful approximate null tests.  

Recall that, as a rule, it is difficult to
reliably predict direct CP violating asymmetries
for exclusive hadronic final states due to 
limited knowledge of
strong phases. Indeed a novel feature of our proposed
tests is that through a judicious
use of U-spin and of inclusive final states a sizable class of direct CP asymmetries can be turned
into precision tests of the SM. 
A stricter null test of the SM is also obtained,
if in \eqref{CPincl} $M^0$ is replaced with a photon, a possibility already discussed in the literature
\cite{Soares:1991te}.

The proposed null tests also have additional
experimental advantages. 
Firstly, since 
direct CP asymmetries are involved no time-dependent measurements
are needed. 
Also, since {\it untagged}
final states are used no separation of $X_d$ from 
$X_s$ is required, rendering PWDs in \eqref{CPincl} rather powerful
null test observables. 
Finally, if $M^0$ is a light meson (for instance $K_S$, $\eta'$) the partial decay widths in the endpoint region where $M^0$ is very energetic, with $E_M\sim M_B/2 +O(\Lambda)$, do not depend on the spectator quark at LO and NLO in $1/m_b$ {\cite{cklz-upcoming}. Thus the PWD for both charged and neutral semiinclusive $B$ meson decays 
\beq
\Gamma(B^-/\bar B^0\to M^0 X_{s+d})-\Gamma(B^+/B^0\to \bar M^0 X_{s+d}),
\eeq
vanish in the endpoint region up to U-spin breaking and corrections of higher order in $1/m_b$. This has a big experimental advantage since
for these decay configurations, of isolated energetic meson $M^0$ and a back-to-back inclusive hadronic jet, only the flavor but not the charge of $B$ meson needs to be tagged.

The remainder of this paper is structured as follows: in Section \ref{sec:SU3limit} we show that \eqref{CPincl} vanishes in the SU(3) limit and then give numerical estimates of the U-spin breaking effects for a few examples in Section \ref{sec:SU3breaking}. Finally, conclusions are gathered in Section \ref{sec:conclsuions}.

%%%%%%%%%%%%%%%%%%%%%%%%%%%%%%%%%
%%%%%%%%%%% II %%%%%%%%%%%%%
%%%%%%%%%%%%%%%%%%%%%%%%%%%%%%%%%
\section{SU(3) limit}\label{sec:SU3limit}
Let us first show that PWD \eqref{CPincl} vanishes in the exact U-spin limit \cite{Soares:1991te,Gronau:2000zy}.
To simplify the notation we take $M^0$ to be
a U-spin singlet $\eta'$ or a $c \bar c$ bound state, while the end result applies also to the
other choices of $M^0$ that were listed above.  Using the decomposition of the $\Delta S=1$ decay width
\beq
\Gamma(B^-\to M^0 X_s^-)=|\lambda_c^{(s)} A_c^s+\lambda_u^{(s)} A_u^s|^2,
\eeq
where $A_{u,c}^s$ denote the terms in the amplitude proportional to the corresponding CKM matrix elements
$\lambda_c^{(s)}=V_{cb} V_{cs}^*\sim \lambda^2$ and $\lambda_u^{(s)}=V_{ub} V_{us}^*\sim \lambda^4$ 
(with $\lambda=\sin\theta_c=0.22$), the corresponding $\Delta S=1$ PWD is
\beq
\begin{split}\label{PWDs}
\Delta \Gamma^s&=\Gamma(B^-\to M^0 X_s^-)-\Gamma(B^+\to M^0 X_s^+)\\
&= -4 J \Im[A_c^{s} A_u^{s*}],
\end{split}
\eeq
with $J=\Im[\lambda_c^{(s)}\lambda_u^{(s)*}]=-\Im[\lambda_c^{(d)}\lambda_u^{(d)*}]$, the Jarlskog invariant. Note that 
$A_{u,c}^s$ are complex since they carry
strong phases. 
Similarly for the $\lambda^2$ suppressed $\Delta S=0$
decay
\beq
\Gamma(B^-\to M^0 X_d^-)=|\lambda_c^{(d)} A_c^d+\lambda_u^{(d)} A_u^d|^2,
\eeq
and
\beq
\begin{split}\label{PWDd}
\Delta \Gamma^d&=\Gamma(B^-\to M^0 X_d^-)-\Gamma(B^+\to M^0 X_d^+)\\
&= 4 J \Im[A_c^{d} A_u^{d*}].
\end{split}
\eeq
The transformation $s\leftrightarrow d$ exchanges $X_s$ and $X_d$ final states, while it has no effect
on $B^\pm$ and $M^0$ states. 
In the limit of exact U-spin thus $A_{u,c}^s=A_{u,c}^d$, giving a vanishing 
PWD in flavor untagged inclusive decay
\beq\label{relationSU3}
\begin{split}
\Delta \Gamma^{s+d}=\Delta \Gamma^s+\Delta \Gamma^d&=4 J\Im[A_c^{d} A_u^{d*}-A_c^{s} A_u^{s*}]=0.\\
\end{split}
\eeq
To the extent that U-spin is a valid symmetry of strong interactions the observable $\Delta \Gamma^{s+d}$ 
constitutes a null test of SM. The breaking can be parameterized completely generally as
\beq\label{SU3break}
\begin{split}
\Delta \Gamma^{s+d}&\equiv\delta_{s\leftrightarrow d} \Delta \Gamma^s ,
\end{split}
\eeq
leading to an expectation for the CP asymmetry of the decay into untagged light flavor 
\beq\label{CPasymmetry}
{\cal A}_{CP}^{s+d}=\frac{\Delta \Gamma^s+\Delta \Gamma^d}{\bar \Gamma^{s+d}+{\Gamma}^{s+d}}
\sim \delta_{s\leftrightarrow d} \frac{\Delta\Gamma^s}{\Gamma^s+\bar\Gamma^s},
\eeq
where in the last relation we have neglected the CKM suppressed $\Gamma^d\sim \lambda^2 \Gamma^s$ decay amplitudes. 
The size of U-spin breaking parameter 
$\delta_{s\leftrightarrow d}$ is channel dependent with an order of magnitude expectation 
$\delta_{s\leftrightarrow d}\sim m_s/\Lambda\sim 0.3$. This is the gain in the theoretical accuracy that one obtains by
summing the $\Delta S=1$ and $\Delta S=0$ PWDs. Summing the two PWDs on the other hand is not expected to reduce
the effect of new physics operators, since unlike SM contributions there is in general no reason for them to give opposite contributions in $\Delta S=1$ and $\Delta S=0$ transitions.

%%%%%%%%%%%%%%%%%%%%%%%%%%%%%%%%%
%%%%%%%%%%% III %%%%%%%%%%%%%
%%%%%%%%%%%%%%%%%%%%%%%%%%%%%%%%%
\section{Concrete examples and the $SU(3)$ breaking}\label{sec:SU3breaking}
We next give several examples of null tests in the semiinclusive hadronic decays. For each of them we also discuss
how reliably we can control the size of SU(3) breaking parameters $\delta_{s\leftrightarrow d}$.

\subsection{PWDs in $B^-\to D^0(\bar D^0) X_{s+d}^-$}
This is a special case, since each of the decays $B^-\to D^0 X_{s}^-$ and $B^-\to \bar D^0 X_{s}^-$ is a pure ``tree" decay with only one CKM structure multiplying the amplitude. This gives vanishing CP asymmetries
\beq\label{BDseparate}
 \Delta \Gamma(B^-\to D^0 X_{s}^-)=\Delta\Gamma(B^-\to \bar D^0 X_{s}^-)=0,
\eeq
and similarly for the $\Delta S=0$ decays $B^-\to D^0(\bar D^0) X_{d}^-$. One thus has trivially
\beq\label{BDcombined}
\Delta \Gamma(B^-\to D^0 X_{s+d}^-)=\Delta \Gamma(B^-\to \bar D^0 X_{s+d}^-)=0
\eeq
without any SU(3) breaking corrections. The nonzero CP
asymmetries arise here only from higher order {\it electroweak} corrections, for instance from a box diagram, giving a CP
asymmetry well below a permil level. This makes either the summed \eqref{BDcombined} or separate $\Delta S=0,1$
CP asymmetries \eqref{BDseparate} clean probes of NP contributions despite the fact that the branching ratios are dominated by the (CKM suppressed) tree level SM transitions.

Note that to measure $\Delta \Gamma(B^-\to D^0 X_{s,d}^-)$ one needs to tag the $D^0$ flavor, for instance using semileptonic $D$ decays or flavor specific $D$ decays. The experimental difficulties in tagging the $\bar D^0$ flavor in 
$\Delta \Gamma(B^-\to D^0 X_{s,d}^-)$ are the same as in $\Delta \Gamma(B^-\to D^0 K^-)$ and
will not be repeated here \cite{Atwood:1996ci}. If alternatively $D^0$ and $\bar D^0$ are decaying to a common final state $f$, one would still have $\Delta \Gamma(B^-\to [D]_f X_{s+d}^-)=0$ in the $SU(3)$ limit, but the SU(3)
breaking corrections are hard to quantify in this case due to a lack of a reliable calculational tool.

We comment in passing that  one also has $ \Delta \Gamma(B^-\to D^0 K^-)=\Delta\Gamma(B^-\to D^0 \pi^-)=0$ up to higher order electroweak corrections, so that these two body decay CP asymmetries can equally be used as null tests of SM. 
Another possibility that avoids $D^0$ flavor tagging is to sum over the $D$ final states. Namely, the decay width for the $B^-\to D K^-$ decay where a sum over 
neutral $D$ meson decays is taken, is an incoherent sum of the decays into $D^0$ and $\bar D^0$, 
$\Gamma(B^-\to D^0 K^-)$+$\Gamma(B^-\to \bar D^0 K^-)$ (and similarly for $B^-\to D \pi^-$ or $B^-\to D X_{s,d}^-$). Each of these decays is a ``tree" decay with only one CKM structure multiplying the amplitude, giving a vanishing CP asymmetry.  

The discussed CP asymmetries can become nonzero in the presence of NP, if the new contributions
have a different weak phase from the SM one and lead to a 
different chiral structure of the effective four-quark operators, giving
a nonzero strong phase difference to the SM contribution (the factorization in $B^-\to D^0 K^-$ indicates that
these phase differences would be $1/m_{b,c}$ suppressed, while in $B^-\to \bar D^0 K^-$ they could be $O(1)$). 
For example, using the recent analysis of SUSY effects on $\gamma$ extraction from $B\to DK$ decays \cite{Khalil:2006zb}, one can conclude that a generic contribution in $R$ parity conserving MSSM that obeys other FCNC constraints
from $B\to X_s \gamma$ and $D^0-\bar D^0$ mixing could lead to a percent level CP asymmetry in $B^-\to D^0(\bar D^0) X_{s}^-$ (depending on the sizes of nonpetrubative strong phase differences between SM and NP operators). Larger
CP asymmetries are possible if gluinos are more massive than quarks or if  $D^0-\bar D^0$ constraints are avoided through partial cancellations between different terms. Similar percent level effects can be expected in  many other extensions of the SM, for instance
in the two-Higgs doublet models where the charged Higgs exchanges can 
lead to enhanced CP asymmetries at the level of several percents~\cite{t2hdm1}.

%%%%%%%%%%%%%%%%%%%%%%%%%%%%%%%%%
%%%%%%%%%%% IV %%%%%%%%%%%%%
%%%%%%%%%%%%%%%%%%%%%%%%%%%%%%%%%
\subsection{Decays into $K X_{s+d}$}
As a next example let us turn to the $B\to K_{S,L} X_{s+d}$ decays. 
We restrict the discussion to the endpoint region 
of the decay phase space with energetic neutral kaon going in the $\bar n=(1,0,0,1)$ direction, 
and the particles in $X_{s,d}^-$ forming an energetic jet with invariant mass 
$p_X^2\sim \Lambda_{\rm QCD} m_B$ going in the $ n=(1,0,0,-1)$ direction. For this kinematic setup SCET 
\cite{Bauer:2000ew,Bauer:2000yr}
offers a theoretical framework \cite{Chay:2006ve}
that will allow us to assess the size of U-spin breaking using already available nonperturbative input from lattice QCD (in the limit of perturbative charming penguins). 
For simplicity we concentrate on the $\Delta S=1$ decay $B^-\to \bar K^0 X_d^-$, while the results will be easy to 
extend
to the $\Delta S=0$ decay $B^-\to K^0 X_s^-$, as well as to the decays involving $K^*$ 
vector mesons. More importantly, the results will also apply to the $\bar B^0\to \bar K^0 X^0, K^0 X_s^0$ decays, since the 
contributions where the spectator ends up in  the energetic $\bar K^0$ meson are $1/m_b^2$ suppressed \cite{cklz-upcoming}.

The relevant part of the 
${\rm SCET}_I$ effective weak Hamiltonian is
\beq\label{HWSCET}
H_W=\frac{2G_F}{\sqrt{2}}\sum_{n,\bar n}
\int\big[d\omega_j\big]_{j=1}^3 c_4^{(s)}(\omega_j)Q_{4s}^{(0)}(\omega_j)+\dots,
\eeq
with the ellipses denoting operators that do not contribute to $B^-\to \bar K^0 X_d^-$. 
The only contributing operator of leading order in $\sqrt{\Lambda/m_B}\sim 0.3$ 
expansion is
\begin{align}\label{QZero}
Q_{4s}^{(0)}&=\sum_q [\bar q_{n, \omega_1}\sla \bar n P_L b_v] [\bar s_{\bar n,\omega_2} \sla n P_L q_{\bar n, \omega_3}].
\end{align}
Here the same notation along with the numbering of the operators 
has been used as in \cite{Bauer:2004tj}. 
Note that the Wilson coefficient $c_4^{(s)}$ in \eqref{HWSCET}
already contains CKM elements \cite{Beneke:1999br,Chay:2003ju,Chay:2006ve}
\beq\label{c4}
c_4^{(s)}(x)=\lambda_u^{(s)} {\cal C}^u_4(x)+\lambda_{c }^{(s)} {\cal C}^c_{4}(x),
\eeq
where to NLO in $\alpha_S(m_b)$ the 
hard kernels at $\mu=m_b$ are $(p=u,c)$ 
\beq\label{tpu}
\begin{split}
{\cal C}_{4}^p(x)=& C_4+\frac{C_3}{N} +\frac{\alpha_S}{4\pi}\frac{C_F}{N}\Big[C_1\Big(\frac{2}{3} -G(s_p)\Big)-\\
&
-\frac{2 C_g}{1-x}\Big]+\cdots,
\end{split}
\eeq
with the ellipses denoting an order of magnitude smaller terms that can be found in Appendix A of \cite{Chay:2006ve}. In the matching of full QCD effective weak Hamiltonian to SCET$_{I}$ weak Hamiltonian
the strong phases are generated at the order $\alpha_S(m_b)$ from the configurations with on-shell intermediate quarks 
carrying a collective momentum $p^2\sim m_b^2$.  The largest contribution to the strong phase comes 
from the tree operator $Q_{1}$  with the $\bar u, u$ or $\bar c, c$ legs contracted, leading to a complex function
($s_c=m_c^2/m_b^2-i \epsilon, s_u=0-i \epsilon$)
\beq\label{Gsp}
G(s_p)=-4 \negthickspace\int_0^1 \negthickspace\negthickspace dz z(1-z) \log \big[s_p-z(1-z)(1-x)\big],
\eeq
with the parameter $x$ denoting the momentum fraction carried by the $s$ quark. This then leads to nonzero CP asymmetries even though only one SCET$_{I}$
operator \eqref{QZero} contributes to the decays considered.

An open problem in the construction of SCET$_{I}$ weak Hamiltonian is the size of 
the long distance contributions coming from intermediate charm quarks annihilating into two collinear quarks, 
where charm quarks are 
in the nonrelativistic QCD regime with small relative velocity. The view of BBNS \cite{Beneke:1999br} 
is that the phase space suppression of the threshold region is strong enough so that nonperturbative contributions are 
subleading, while Bauer et al. \cite{Bauer:2004tj} argue that the phase space suppression is not
effective as numerically $2 m_c/m_b\sim O(1)$. For semiinclusive hadronic decays the factorization of charming penguin
contributions into soft and collinear parts has been shown in \cite{Chay:2006ve}. We will first proceed as though
charm quarks can be perturbatively integrated out leading to $C_1$ term in \eqref{tpu} for $p=c$. The effect of nonperturbative charming penguins  will then be discussed  at the end of present subsection.

An important observation in deriving the expression for the partial decay width is that the $\bar n$ and $n$ parts of the operator $Q_{4s}^{(0)}$ in \eqref{QZero}
decouple from each other at leading order in $1/m_b$. Making redefinitions $q_{n, \bar n}\to Y_{n, \bar n} q_{n, \bar n}$
and $A_{n, \bar n}\to Y_{n, \bar n} A_{n, \bar n} Y_{n, \bar n}^\dagger$ with $Y_{\{n, \bar n\}}$ a Wilson line 
of ultrasoft gluons $\{n,\bar n\}\cdot A_{\rm us}$, the ultrasoft gluons decouple from collinear fields 
both in the leading order SCET Lagrangian as well as in $Q_{4s}^{(0)}$, where now the ultrasoft Wilson lines multiply only the 
$b_v$ fields, $Y_n^\dagger b_v$ 
\cite{Bauer:2001yt,Chay:2003ju}. At leading order in $1/m_b$ operator $Q_{4s}^{(0)}$ thus factorizes to
\beq
Q_n^q(\omega_1)=[\bar q_{n, \omega_1}\sla \bar n P_L Y_n^\dagger b_v],
\eeq
and the remaining $ Q_{4s,q}^{\bar n}$ pieces that do not talk to each other. It is this factorization that makes
predictions of semi-inclusive decays $B^-\to \bar K^0 X_d^-$ in SCET$_{\rm I}$ region possible. 
The decay amplitude then factorizes into matrix elements of operators in $n$ and 
$\bar n$ directions
\beq\label{FactorAmpl}
\begin{split}
\langle X_d^- \bar K^0| &H_W|B^-\rangle = \frac{2G_F}{\sqrt{2}}\sum_{q} \int\negthickspace\big[d\omega_j\big]_{j=1}^3 c_4^{(s)}(\omega_j) \\
\times &\Big[\langle X_d^-| Q_{n}^{q}(\omega_1)|B^-\rangle \langle \bar K^0| Q_{4s,q}^{\bar n}(\omega_j)|0\rangle\\
&+\langle \bar K^0| Q_{n}^{q}(\omega_1)|B^-\rangle \langle X_d^-| Q_{4s,q}^{\bar n}(\omega_j)|0\rangle\Big].
\end{split}
\eeq
In the decay of $B^-$ the spectator $\bar u$ quark cannot end up in $\bar K^0$, so that 
$\langle \bar K^0| Q_{n}^{q}(\omega_1) |B^-\rangle=0$ and the
last term in \eqref{FactorAmpl} vanishes. In the sum thus only $q=d$ contribution in the first term is nonzero.

Since $\bar K^0$ decouples to leading order from the rest of the amplitude, one can calculate the inclusive decay width 
$\bar \Gamma_{K}^{(s)}\equiv\Gamma(B^-\to \bar K^0 X_d^-)$ \footnote{Note that the superscript $\bar \Gamma_{K}^{(s)}$ denotes that this is a $\Delta S=1$ decay. It does not
denote the net strangeness content of the inclusive jet $X_d$.} following the same steps as in the SCET calculation of 
$\Gamma(B\to X_s \gamma)$ in the endpoint region \cite{Bauer:2001yt} (with more details given in \cite{Chay:2006ve}). 
The inclusive decay width is simply a product of two terms, one 
coming from $\bar n$, the other from $n$ 
operators
\beq
\begin{split}\label{dG}
\frac{d\bar \Gamma_K^{(s)}}{d E_K}=&\frac{G_F^2}{\pi}
 \;E_K^3 \left|\int_0^1 dx f_K\phi_K(x) c_4^{(s)}(x)\right|^2 \times\\
&\int_{2E_K-m_b}^{\bar \Lambda} dk^+ S(k^+) J(k^++m_b-2E_K).
\end{split}
\eeq
Multiplying \eqref{dG} by an extra factor of $1/2$ gives $\Gamma(B^-\to K_{S,L} X_d^-)$. The 
$\Gamma(B^+\to K^0 X_d^+)\equiv \Gamma_{K}^{(s)}$ decay width is obtained by changing in Eq. \eqref{dG}
$\lambda_u^{(s)}\to \lambda_u^{(s)*}$ 
in $c_4^{(s)}(x)$. 
The $B$ meson shape function $S(k^+)$ and the perturbatively calculable jet function $J(k^+)$
are exactly the same as the ones found in the decay $B\to X_q\gamma$, 
with their definitions given in \cite{Bauer:2001yt}. 
For decays with $K^*$ vector meson only the decay into longitudinal polarization state is nonzero at leading order in 
$1/m_b$. The decay width
is then obtained from \eqref{dG} by replacing 
the kaon decay constant $f_K\to f_{ \parallel K^*}$ and the light-cone distribution amplitude 
$\phi_K\to \phi_{\parallel K^* }$. 

Normalizing the difference of the decay widths $\Delta\Gamma^{(s)}_{K}\equiv\Gamma(B^+\to K^0 X_d^+)-\Gamma(B^-\to \bar K^0 X_d^-)=
\Gamma^{(s)}_{K}-\bar \Gamma^{(s)}_{K}$ with their sum, the shape and jet functions drop out and so does the
dependence on $E_K$. To first order in $\lambda^2$ suppressed terms thus, in SCET$_{I}$ region
\beq
\begin{split}\label{deltadG}
{\cal A}_{CP}^{(s)}=\frac{\bar \Gamma^{(s)}_{K}-\Gamma^{(s)}_{K}}{\Gamma^{(s)}_{K}+\bar\Gamma^{(s)}_{K}}=&
-2 J \frac{\Im \left[{\cal T}_{K,c}^{(s)}{\cal T}_{K,u}^{(s)*}\right]}{|\lambda_c^{(s)} {\cal T}_{K,c}^{(s)}|^2},
\end{split}
\eeq
where the hard kernels ${\cal T}_{K,p}^{(s)}$ are 
\beq\label{TK}
{\cal T}_{K,p}^{(s)}=f_K\int dx \phi_K(x) {\cal C}_4^p(x),
\eeq
with ${\cal C}_4^{u,c}(x)$ given in \eqref{c4}, \eqref{tpu}. If we can neglect nonperturbative charming penguin contributions, the CP asymmetry \eqref{deltadG} depends only on one nonperturbative function, the kaon LCDA $\phi_K(x)$.
In the numerical results we will use a recent lattice QCD determination of the first coefficient in the Gegenbauer expansion of  $\phi_K(x)$ which at $\mu=2.0$ GeV is $a_1^K=0.055\pm0.005$ \cite{Boyle:2006pw}. This value is in agreement
with a recent QCD sum rule analysis \cite{Ball:2006wn},  that 
gives at $\mu=2$ GeV:  $a_1^K=0.05\pm0.03$, $a_2^K=0.23\pm0.12$ (we use this value for $a_2^K $ in the numerical analysis but conservatively double the errors).
Using the values of CKM elements from \cite{Charles:2004jd} and running the SCET$_{\rm I}$ Wilson coefficients \eqref{c4}, \eqref{tpu} to $\mu=2.0$ GeV using NLL RG equations \cite{Chay:2006ve}, we obtain
\beq\label{numdG}
\begin{split}
{\cal A}_{CP}^{(s)}&=
(0.27\pm0.05)\times ({2 J}/{|\lambda_c^{(s)}|^2}) \\
&=(1.0\pm0.2)\cdot 10^{-2},
\end{split}
\eeq
where the errors  reflect only the errors on Gegenbauer coefficients $a_{1,2}^K$, with the error on $a_{2}^K$ dominating.

In $\phi_K(x)$ parameter $x$ denotes the fraction of kaon momentum carried by the
strange quark, while in the hard kernels ${\cal T}_{K,p}^{(s)}(x)$ it denotes the momentum carried by quark (antiquark) in the $K$ meson 
starting with a $\bar B$ ($B$) initial meson. Thus the difference of the $\Delta S=0$ decay widths 
$\Delta \Gamma^{(d)}_{K}\equiv \Gamma(B^-\to \bar K^0 X_s^-)-\Gamma(B^+\to K^0 X_s^+)$ 
is obtained from \eqref{deltadG} by replacing $J\to -J$ and $\phi_K(x)\to \phi_K(1-x)$. It is therefore useful to decompose
$\phi_K(x)$ into functions $\phi_K^{\pm}(x)$ that are even and odd  under the $x\to 1-x$ exchange
\beq
\phi_K(x)=\phi_K^+(x)+\phi_K^-(x), \quad \phi_K^\pm(x)=\pm \phi_K^\pm(1-x).
\eeq
In the Gegenbauer polynomial expansion $\phi_K^+$ ($\phi_K^-$) receives contributions only from 
even (odd) Gegenbauer polynomials. 
Defining similarly the corresponding hard kernels
\beq
{\cal T}_{u,c}^\pm=f_K \int dx\, \phi_K^\pm (x)\,{\cal C}_4^{u,c}(x),
\eeq
we get for the sum of the CP asymmetries (to the first order in the CKM suppressed terms)
\beq\label{sumdG}
\begin{split}
{\cal A}_{CP}^{d+s}=&\big({ \Delta\Gamma^{(s)}_{K}+\Delta\Gamma^{(d)}_{K}}\big)/\big({ \Gamma^{d+s}_{K}+\bar\Gamma^{d+s}_{K}}\big)=\\
=&-4J {\Im \left[{\cal T}_{c}^{+}{\cal T}_{u}^{-*}+{\cal T}_{c}^{-}{\cal T}_{u}^{+*}\right]}/
{|\lambda_c^{(s)}{\cal T}_{c}|^2},
\end{split}
\eeq
which using the same input values as in \eqref{numdG} gives
\beq
\begin{split}
{\cal A}_{CP}^{d+s}
&=(7.6\pm 6.4)\cdot 10^{-2}\times {4J}/{|\lambda_c^{(s)}|^2}\\
&=(0.28\pm0.23)\cdot 10^{-2},
\end{split}
\eeq
where again the errors only show the dependence on kaon LCDA. Note that in the limit of exact U-spin ${\cal T}_{c}^{-}={\cal T}_{u}^{-}=0$ and 
therefore ${\cal A}_{CP}^{d+s}=0$.  Eq. \eqref{sumdG} encompasses the U-spin breaking effect due to asymmetric $\phi_K(x)$, and gives a reasonable estimate for the size of  $|{\cal A}_{CP}^{d+s}|$ also in the case of nonperturbative charming penguins as we discuss next.
The remaining SU(3) breaking due to $m_s$ suppressed SCET operators lead to additional jet functions of order
$m_s^2/\Lambda m_b$ and can be safely neglected \cite{Chay:2005ck}.

To LO in $\Lambda/m_c$ the nonperturbative charming penguin contributions in semiinclusive hadronic decays factorize
into an $\bar n$ collinear factor that depends on meson $M$ and a universal convolution ${\cal F}_{cc}$ of soft ``charm shape function" and an $n$-collinear jet function as shown in \cite{Chay:2006ve}. Normalizing the decay width to the $B\to X_s\gamma$ decay, so that 
the jet function and the shape function in \eqref{dG} cancel for the perturbative hard kernels, we find that in the endpoint region
\beq
\begin{split}\label{ratioG}
&\frac{d\Gamma(B^-\to \bar K^0 X_d^- )}{d\Gamma(B\to X_s \gamma)}=\frac{2\pi^3}{\alpha m_b^2}
\frac{1}{|\lambda_t^{(s)}C_\gamma(c_9^{\rm eff}+1/2 c_{12}^{\rm eff})|^2}\times\\
&\qquad\quad\times \Big\{
\Big|\sum_{p=u,c}\lambda_p^{(s)}{\cal T}_{K,p}^{(s)}\Big|^2+f_K^2|\lambda_c^{(s)}|^2\overline F_{cc}\\
&\qquad\qquad+2 \Re\Big[f_K\lambda_c^{(s)}\bar f_{cc}\Big(\sum_{p=u,c}\lambda_p^{(s)}{\cal T}_{K,p}^{(s)}\Big)^*\Big] 
\Big\}
,
\end{split}
\eeq
where one sets $E_\gamma=E_K$. The SCET Wilson coefficients are
$c_9^{\rm eff}=1$, $c_{12}^{\rm eff}=0$ at LO with NLO calculated in \cite{Bauer:2000yr}, while $C_\gamma$ is given 
e.g. in Eq. (13) of  \cite{Lee:2004ja}. The complex parameter $\bar f_{cc}$ that describes the interference of nonperturbative charming penguin with the perturbative hard kernels is related to the  soft charm shape function ${\cal F}_{cc}$ defined in \cite{Chay:2006ve}
\beq\label{fcc}
\bar f_{cc}=\frac{\alpha_S(2 m_c)}{m_b}\frac{{\cal F}_{cc}\;\phi_M\left(1-\frac{2m_c^2}{E_M m_b}\right)}{  S(k^+) \otimes_{k^+} J(k^++m_b-2E_K)},
\eeq
where $\otimes_{k^+}$ denotes the integration over $ k^+\in [{2E_K-m_b},{\bar \Lambda}]$. The parameter ${\cal F}_{cc}$ is
universal for any $\Gamma(B\to MX)$ up to $O(\Lambda/m_c)$ corrections, while $\bar f_{cc}$ depends on meson $M$'s LCDA $\phi_M$. The positive real parameter $\overline F_{cc}$ in \eqref{ratioG} on the other hand
describes the square of nonperturbative charming penguin contributions. As a rule of thumb we can thus take $\bar f_{cc}^2 \sim \overline F_{cc}$. Similarly to $\bar f_{cc}$ the parameter $\overline F_{cc}$ depends on meson $M$ through $\phi_M^2\left(1-{2m_c^2}/{E_M m_b}\right)$. If hard kernels dominate the amplitudes, the term with $\bar f_{cc}$  in \eqref{ratioG} is subleading, while $\overline F_{cc}$ term is even more suppressed and can be neglected as was done in \cite{Chay:2006ve}. It should, however, be kept in penguin dominated modes. A prediction for $\Delta S=0$ decay width $\Gamma(B^-\to  K^0 X_s^- )$ in the presence of nonperturbative charming penguins is obtained from \eqref{ratioG} by making a replacement $s\to d$, where ${\cal T}_{K,p}^{(d)}$ is obtained from
\eqref{TK} through a replacement $\phi_K(x)\to \phi_K(1-x)$. The $B^+$ decay widths are obtained by making a replacement $\lambda_p^{(q)}\to 
\lambda_p^{(q)*}$.

The results derived in this subsection are valid also for $B^0\to K_S X_{s,d}^0$ semiinclusive hadronic decays in the endpoint region up to power suppressed corrections. These arise from the second term in \eqref{FactorAmpl} describing the spectator interactions and lead to $1/m_b^2$ correction to the decay widths. Up to these corrections all results, including numerical ones, are the same for charged and neutral $B\to K_S X_{s,d}$ decays.

We can use this fact to determine the charming penguin parameters from the presently available experimental data. 
Recently the first measurement of 
$B\to K^0 X$ 
branching ratio was reported by BaBar \cite{Aubert:2006an}
\beq
\begin{split}
Br(B\to K^0 X)&=(154^{+55+55}_{-48-41})\cdot 10^{-6},
\end{split}
\eeq
where the lower cut on the $K$ momentum of $2.34$ GeV in the $B$ rest frame was used. Normalizing to the $B\to X_s\gamma$
branching ratio with the same photon momentum cut one has \cite{HFAGSummer2006,Buchmuller:2005zv}
\beq\label{ratioKX}
\frac{Br(B\to K^0 X)}{Br(B\to X_s\gamma)}=0.89\pm0.43,
\eeq
The prediction for this ratio is given in \eqref{ratioG} once it is $CP$ averaged (alternatively, to accuracy we are working one can neglect $\lambda_u$ suppressed terms). It depends on three nonperturbative parameters, $\overline {F}_{cc}$ and magnitude and phase of $\bar{f}_{cc}$. At present there is not enough experimental information to determine all three of them. Quite generally one expect $\bar |f_{cc}|^2 \sim \overline F_{cc}$. As a starting point, we take this relation to be exact, which leads to $\sqrt{\overline F_{cc}}=(8.9\pm6.6)\cdot 10^{-2}$, where the error is a sum of experimental error and the variation of $\arg(\bar f_{cc})\in [0,2\pi)$. This value of $\sqrt{\overline F_{cc}}$ is about a factor of $4\pm 3$ larger then the perturbative prediction for the charming penguin 
\eqref{tpu} (with  $\sqrt{\overline F_{cc}}=0$ corresponding to purely perturbative charming penguin). Experimentally, there is therefore 
a possible indication for sizable nonperturbative charming penguin, but the data are at present also consistent with $\sqrt{\overline F_{cc}}=0$ at a little above $1\sigma$. For instance, neglecting nonperturbative charming penguins gives $0.19\pm0.04$ for the ratio in \eqref{ratioKX}.

For nonzero nonperturbative charming penguin contributions the CP asymmetry ${\cal A}_{CP}^{d+s}$ is governed by the size of SU(3) breaking in the charming penguin. Taking a $30\%$ SU(3) breaking with $\bar f_{cc}^2 \sim \overline F_{cc}$, gives 
\beq
{\cal A}_{CP}^{d+s}\in[-0.6\%,0.9\%],
\eeq
to be compared with 
\beq
{\cal A}_{CP}^{s}\in[-2.3\%,2.3\%],
\eeq
that is obtained for the same set of input parameters. This illustrates the benefit of using combined CP asymmetry ${\cal A}_{CP}^{d+s}$, where the theoretical uncertainties are reduced in two ways: (i) the central value is reduced, since ${\cal A}_{CP}^{d+s}$ vanishes in $SU(3)$ limit, while ${\cal A}_{CP}^{s}$ does not, and (ii) the error on the prediction is reduced. To understand how this happens, 
let us look at the contribution of nonperturbative charming penguins to the rate asymmetries
\begin{align}
\Delta \Gamma^{(s)}_K&\propto \Im[({\cal T}_u^++{\cal T}_u^-)(\bar f_{cc}^{K})^*],\\
\Delta \Gamma^{(s+d)}_K&\propto \Im[{\cal T}_u^+(\bar f_{cc}^K-\bar f_{cc}^{\overline K})^*+{\cal T}_u^-(\bar f_{cc}^K+\bar f_{cc}^{\overline K})^*],
\end{align}
where we have explicitly denoted the dependence of $\bar f_{cc}$ on $M=K,\bar K$ (cf. Eq. \eqref{fcc}).
In the $SU(3)$ limit $\bar f_{cc}^K=\bar f_{cc}^{\overline K}$ and ${\cal T}_u^-=0$, so that the contribution of
nonperturbative charming penguins to $\Delta \Gamma^{(s+d)}_K$ vanishes as expected. Furthermore, if in the future  $\bar f_{cc}^K$ and $\bar f_{cc}^{\overline K}$ are determined from some other decay modes such as semiinclusive decays involving charged kaons, the resulting error on the prediction of $\Delta \Gamma^{(s+d)}_K$ will be smaller then for $\Delta \Gamma^{(s)}_K$ since the error in the difference $\bar f_{cc}^K-\bar f_{cc}^{\overline K}$ partially cancels, while the error on $(\bar f_{cc}^K+\bar f_{cc}^{\overline K})^*$ comes multiplied by the $SU(3)$ breaking factor ${\cal T}_u^-$.

Finally, we also give the results for $B\to (K^{*0}+\bar K^{*0}) X_{s,d}$ decays that can be trivially obtained from the above results with the replacement $\phi_K(x)\to \phi_{K^*\parallel}(x)$. Using $a_1^{K^*}=0.08\pm0.13$, $a_2^{K^*}=0.07\pm0.08$ at $\mu=2.0$ GeV obtained by conservatively doubling the errors of \cite{Ball:2004rg}, we get
\beq
\begin{split}
{\cal A}_{CP, K^*}^{(s)}&=
(0.24\pm0.02)\times ({2 J}/{|\lambda_c^{(s)}|^2}) \\
&=(0.86\pm0.07)\cdot 10^{-2},
\end{split}
\eeq
and
\beq
\begin{split}
{\cal A}_{CP, K^*}^{d+s}
&=(3.2\pm 2.6)\cdot 10^{-2}\times {4J}/{|\lambda_c^{(s)}|^2}\\
&=(0.12\pm0.10)\cdot 10^{-2},
\end{split}
\eeq
and for the ratio of decay widths
\beq
\frac{d\Gamma(B^-\to \bar K^{0*} X_d^- )}{d\Gamma(B\to X_s \gamma)}=0.33\pm0.11,
\eeq
where as before the errors are only due to error on $K^*$ LCDA. These predictions do not include effects of nonperturbative charming penguins. Using the determination of $\bar F_{cc}$ from Eq. \eqref{ratioKX} and taking a $30\%$ SU(3) breaking with $\bar f_{cc}^2 \sim \overline F_{cc}$, gives 
\begin{align}
&{\cal A}_{CP,K^*}^{d+s}\in[-0.6\%,0.8\%],\\
&{\cal A}_{CP,K^*}^{s}\in[-2.1\%,2.1\%],
\end{align}
and
\beq
\frac{d\Gamma(B^-\to \bar K^{0*} X_d^- )}{d\Gamma(B\to X_s \gamma)}=1.66\pm0.86,
\eeq
where the $\Lambda/m_c$ suppressed contributions from decays into transverselly polarized $K^*$ have been neglected.

\subsection{$B^-\to \eta'X_{s+d}^-$}
We next turn to the case of $B^-\to \eta'X_{s+d}^-$ decay, by
first showing that the U-spin breaking $\delta_{s\leftrightarrow d}$ is still linear
in $m_s/\Lambda$. In particular the $\eta-\eta'$ mixing does not introduce anomalously large breakings. We
use the FKS mixing scheme \cite{Feldmann:1998vh} in
which the mass eigenstates
$\eta$, $\eta'$ are related to the flavor basis through
%\beq
$\eta=\eta_q \cos \varphi -\eta_s \sin \varphi,$ % \quad
and $\eta'=\eta_q \sin \varphi +\eta_s \cos \varphi, $
%\eeq
with $\varphi=(39.3\pm1.0)^\circ$ and $\eta_q=(\eta_u+\eta_d)/\sqrt{2}$. 
We start by rewriting \eqref{relationSU3}
\beq
\begin{split}
\Delta \Gamma(\eta' X^\pm_{s+d})=-4 J\Im[\Delta A_c A_u^{s*}+A_c^{d} \Delta A_u^{*}],\\
\end{split}
\eeq
where the flavor breaking difference $\Delta A_c=A_c^s-A_c^d$ is 
\beq\label{DeltaAc}
\begin{split}
\Delta A_c&=\frac{\sin\varphi}{\sqrt 2} \Big\{\big[ A_c(\eta_uX_s^-)+A_c(\eta_dX_s^-)+A_c(\eta_sX_s^-)\big]
\\
-\big[s&\leftrightarrow d\big]\Big\}+\Big(\cos\varphi-\frac{\sin\varphi}{\sqrt 2} \Big)\Big(A_c(\eta_sX_s^-)-A_c(\eta_sX_d^-)\Big),
\end{split}
\eeq
and similarly for $\Delta A_u=A_u^s-A_u^d$. 
Here $A_c(\eta_qX_s^-)$ denotes a term in the amplitude  
due to a $q\bar q$ part of $\eta'$ wave function. 
The terms in the curly brackets in \eqref{DeltaAc} cancel in the limit of exact $s\leftrightarrow d$ symmetry.
The difference $A_c(\eta_sX_s^-)-A_c(\eta_sX_d^-)$ 
in the last term 
 on the contrary, does
not vanish in the exact U-spin limit (even though there is a partial cancellation). However, the term multiplying it, 
$\cos\varphi-\frac{\sin\varphi}{\sqrt 2}=0.33$, makes its size a typical SU(3) breaking effect, and would
vanish for SU(3) singlet $\eta'$ since then $\tan\varphi=\sqrt 2$. Thus
the corresponding $\delta_{s\leftrightarrow d}$ is of typical size, $O(m_s/\Lambda)$. 

A more quantitative analysis can be made in the endpoint region of the inclusive decay using SCET$_{\rm I}$ in the same way as in the previous subsection. Neglecting the $1/m_b^2$ suppressed spectator interactions \cite{cklz-upcoming} and the $\alpha_S^2(m_b)$ contribution from the gluonic operator $Q_{gs}^{(0)}$ \cite{Williamson:2006hb} one finds for the hard kernels ($p=u,c$, while $\otimes$ denotes a convolution over $x$)
\begin{align}
\begin{split}
{\cal T}_{\eta',p}^{(s)}&=f_{\eta_s}\cos\varphi \;\phi_{\eta_s}\otimes \big({\cal C}_4^p+{\cal C}_5-{\cal C}_6\big)\\
&+f_{\eta_q} \frac{\sin \varphi}{\sqrt{2}}\; \phi_{\eta_q}\otimes\big({\cal C}_2^p-{\cal C}_3+2{\cal C}_5-2{\cal C}_6\big),
\end{split}
\\
\begin{split}
{\cal T}_{\eta',p}^{(d)}&=f_{\eta_s}\cos\varphi \;\phi_{\eta_s}\otimes \big({\cal C}_5-{\cal C}_6\big)\\
&+f_{\eta_q} \frac{\sin \varphi}{\sqrt{2}}\; \phi_{\eta_q}\otimes\big({\cal C}_2^p-{\cal C}_3+{\cal C}_{4}^p+2{\cal C}_5-2{\cal C}_6\big),
\end{split}
\end{align}
in terms of which the CP asymmetries are (neglecting the CKM suppressed terms)
\beq
\begin{split}
&{\cal A}_{CP,\eta'}^{(s)}=\\
&\frac{-2J}{|\lambda_c^{(s)}|^2}{\Im\Big[({\cal T}_{\eta',c}^{(s)}+\cos\varphi f_{\eta_s}\bar f_{cc}){\cal T}_{\eta',u}^{(s)*}\Big]}D^{-1},
\end{split}
\eeq
and
\beq
\begin{split}
{\cal A}_{CP,\eta'}^{(s+d)}=&\frac{-2J}{|\lambda_c^{(s)}|^2}\Big\{\Im\big[({\cal T}_{\eta',c}^{(s)}+\cos\varphi f_{\eta_s}\bar f_{cc}){\cal T}_{\eta',u}^{(s)*}\big]\\
&-\Im\big[({\cal T}_{\eta',c}^{(d)}+\frac{\sin\varphi}{\sqrt2} f_{\eta_q}\bar f_{cc}){\cal T}_{\eta',u}^{(d)*}\big]
\Big\}D^{-1},
\end{split}
\eeq
with
\beq
D=
\Big[|{\cal T}_{\eta',c}^{(s)}|^2+2 \cos\varphi f_{\eta_s}\Re ({\cal T}_{\eta',c}^{(s)} \bar f_{cc})+\overline F_{cc} (\cos\varphi f_{\eta_s})^2\Big].
\eeq
Here the complex parameter $\bar f_{cc}$ and the real positive parameter $\overline F_{cc}$ parameterize the charming penguin contributions in the same way as described in the previous subsection. They can be constrained using the measurements of BaBar \cite{Aubert:2004eq} and CLEO \cite{Bonvicini:2003aw} of $B\to \eta'X_s$ branching ratio. Combining the two measurements gives
\beq
Br(B\to \eta'X_s)=(420\pm94)\cdot 10^{-6},
\eeq
for a lower cut on $\eta'$ energy of $E_{\eta'}>2.218$ GeV. Normalizing to the $B\to X_s \gamma$ branching ratio with  $E_{\gamma}>2.218$ GeV \cite{HFAGSummer2006}
\beq
\frac{Br(B\to \eta'X_s)}{B\to X_s \gamma}=1.74\pm0.42,
\eeq
we can use the expression
\beq
\begin{split}
&\frac{d\Gamma(B^-\to \bar \eta'X_s)}{d\Gamma(B\to X_s \gamma)}=\frac{2\pi^3}{\alpha m_b^2}
\frac{D}{|\lambda_t^{(s)}C_\gamma(c_9^{\rm eff}+1/2 c_{12}^{\rm eff})|^2}
,
\end{split}
\eeq
to constrain $\overline F_{cc}$, while bounds on $\bar f_{cc}$ obtained in this way are very loose. Naively one expects $\overline F_{cc}\sim |f_{cc}|^2$. If this is satisfied, then  $\overline F_{cc}$ dominates in $Br(B\to \eta'X_s)$ leading to a determination $\sqrt{\overline F_{cc}}=0.15\pm0.03$, where the error is a combination of experimental one and due to a variation of $\arg(\bar f_{cc})$ in the determination. This corresponds to a nonperturbative charming penguin, which is about $5-8$ times larger than the perturbative contribution. Whether this is the correct interpretation of the enhancement of $Br(B\to \eta'X_s)$ over the perturbative prediction should be clarified once other 
semiinclusive hadronic decays are measured. For instance,  the parameters $\sqrt{\overline F_{cc}}$ determined in $Br(B\to \eta'X_s)$ and $Br(B\to K^0 X)$ should be the same up to corrections of order $\Lambda/m_c$.  

Using $\overline F_{cc}\sim |f_{cc}|^2$ together with the hard kernels 
calculated using NLO matching at $\mu\sim m_b$ with NLL running to $\mu=2.0$ GeV, setting $f_{\eta_q}=140\pm 3$~MeV, $f_{\eta_s}=176\pm8$~MeV \cite{Feldmann:1998vh} and taking $\phi_{\eta_q}(x)=\phi_{\eta_s}(x)=\phi_\pi(x)$ in the lack of better information, while varying the phase $\arg(\bar f_{cc})\in[0,2\pi)$, we obtain 
\beq
{\cal A}_{CP,\eta'}^{(s)}\in[-1.7\%,1.7\%],
\eeq
and
\beq
{\cal A}_{CP,\eta'}^{(s+d)}\in[-1.2\%,0.9\%].
\eeq
 This is in agreement with a result for 
CP asymmetry of this mode  
${\cal A}_{CP,\eta'}^{(s)}\sim 1\%$ from \cite{Cheng:2001nj}.

In addition to the null tests discussed above there are also other null test that one could consider. For instance neglecting annihilation diagrams
also neutral decay $\bar B^0\to \pi^+ X_{s+d}^-$ has vanishing PWD \eqref{CPincl} in U-spin limit, 
 see e.g. 
 \cite{He:2002ae}. Another interesting case is $\phi X_{s+d}^-$.  Since $\phi$ is not a U-spin singlet
the PWD \eqref{CPincl} does not vanish in the exact U-spin limit. 
Nevertheless, in the SM
this decay is penguin dominated with
very small direct CP asymmetry $\approx 1\%$ \cite{Cheng:2001nj}
providing a valuable probe of NP.

\section{Conclusions}\label{sec:conclsuions}
In light of B-factories' results
it is becoming increasingly clear that deviations from the CKM-
paradigm due to NP are likely to be small.
Therefore null tests of the SM can be very valuable in search
of NP. Bearing that in mind, we are proposing a new class
of null tests involving
CP asymmetries
of
untagged, semi-inclusive decays, $B \to M^0 X_{s + d}$
where $M^0$ is either a U-spin singlet (for instance $D^0$ or $c\bar c$ bound state) or a meson that is related to
its antiparticle through a U-spin rotation. In general the CP asymmetries vanish in U-spin limit only for charged $B$ decays. However, if $M^0$ is an energetic light meson such as $K_{S,L}$, $\eta'$ or $K^*$ (taken together with the decay to its CP conjugate $\bar K^*$), the decaying $B$ can be taken to be either charged or neutral up to $1/m_b^2$ corrections. In the examples discussed  we 
showed that 
these CP asymmetries
are very small, $<1\%$. 
Experimentally, to perform a completely inclusive measurement for these decays the flavor but not the charge of
decaying $B$ meson needs to be tagged. Recently the first measurement of  $B \to K^0 X$ branching ratio was performed 
by BaBar using fully reconstructed $B$ decays \cite{Aubert:2006an}, suggesting that
the proposed  observables are experimentally measurable in practice. \\[2mm] 

\noindent
{\bf Acknowledgements}:

JZ wishes to thank Junegone Chay, Chul Kim and Adam Leibovich for collaboration and enlightening discussions regarding hadronic semiinclusive $B$ decays. We thank H.~Flacher for providing us with the extrapolation factor.
This research was supported in part by DOE contract 
Nos.DE-FG02-04ER41291(BNL), 
DOE-ER-40682-143(CMU) and DEAC02-6CH03000(CMU).

%%%%%%%%%%%%%%%%%%%%%%%%%%%%%%%%%
%%%%%%%%%%%%% bib %%%%%%%%%%%%%
%%%%%%%%%%%%%%%%%%%%%%%%%%%%%%%%%

\end{document}